\documentclass[aps,pre,twocolumn,float]{revtex4}
\usepackage{amsmath,bm,epsfig}
\let \nn  \nonumber

\let\*\cdot
\def\<{\left\langle} \def\>{\right\rangle} \def\({\left(} \def\){\right)}
 \let\~\widetilde \let\^\widehat 

\def\be{\begin{equation}}\def\ee{\end{equation}}
\def\bea{\begin{eqnarray}}\def\eea{\end{eqnarray}}
\def\bse{\begin{subequations}}\def\ese{\end{subequations}}
\newcommand{\BE}[1]{\begin{equation}\label{#1}}
\newcommand{\BEA}[1]{\begin{eqnarray}\label{#1}}
\newcommand{\BSE}[1]{\begin{subequations}\label{#1}}

\let \nn  \nonumber
\usepackage{pstricks}
\usepackage{pst-node}
\usepackage[ansinew]{inputenc}
\usepackage{amssymb,amsmath}

\def\BSE{\begin{subequations}}\def\ESE{\end{subequations}}
\let \= \equiv

\def\a{\alpha}

\def\g{\gamma}

\def\o{\omega}

\def\be{\begin{equation}}       \def\ba{\begin{array}}

\def\ee{\end{equation}}         \def\ea{\end{array}}

\def\bea {\begin{eqnarray}}      \def\eea {\end{eqnarray}}

\def\bean{\begin{eqnarray*}}    \def\eean{\end{eqnarray*}}

\def\const {\mathop{\rm const}\nolimits}

\def\RA {\ \Rightarrow\ }

\def\<{\langle} \def\({\left(}  \def\>{\rangle} \def\){\right)}

\newtheorem{exi}{Example}

\begin{document}

\title{A novel model of wave turbulence}
\author{Elena Kartashova$^{\dag},$}
 \email{Elena.Kartaschova@jku.at}
  \affiliation{$^{\dag}$ Institute for Analysis, J. Kepler University, Linz, A-4040 Austria}  
  \affiliation{$^*$ Kavli Institute for Theoretical Physics, University of California, Santa Barbara, CA 93106, USA}

   \begin{abstract}
A novel D-model of wave turbulence is presented which allows to reproduce in a single frame various nonlinear wave phenomena such as intermittency, formation and direction of  energy cascades, possible growth of nonlinearity due to direct energy cascades, etc. depending on  the initial state. No statistical assumptions are used, all effects are due to the behavior of distinct modes. Classical energy spectra $E_{\o} \sim \o^{-\nu}, \ \nu= \const >0,$ for dispersion function of the form $ \o \sim k^{\a}, \ \a>0$ are obtained as a particular case of a more general form of energy spectra: $E_{\o} \sim \o^{-\nu}, \ 2+\a^{-1}\le \nu \le 2(2+\a^{-1})$ where magnitude of $\nu$ is defined by the parameters of the initial excitation. D-model is a generic model which can be expanded into a hierarchy of more refined models including dissipation, forcing, etc.  D-model can be applied to the experimental and theoretical study of numerous wave turbulent systems appearing in hydrodynamics, nonlinear optics, electrodynamics, convection theory, etc.
\end{abstract}


\maketitle

\textbf{1. Introduction.}

\noindent Stationary energy spectra in the statistical theory of wave turbulence follow the celebrated Kolomogorov-Zakharov (KZ) law $E_k\sim k^{-\nu}$, \cite{ZLF92}, with dispersion relation $\omega\sim k^\alpha$, $\nu, \alpha >0$ being constants, and $k$ the wave number. These  spectra are obtained under certain  assumptions -- like the smallness of non-linearity, infinite boxes, existence of an inertial range $(k_1<k<k_2)$, local interaction in $k$-space etc. -- in close analogy to ordinary turbulence by replacing vortices with waves.

Attempts to verify this theory led, however, to controversial results (for a review see \cite{NR11}), among them in several cases the lack of the generation of the energy cascade and instead the production of regular wave patterns, \cite{HHS03}. Moreover, if a cascade occurs, its spectrum consists of two distinct parts: a discrete spectrum and a continuous spectrum. The former, and frequently even the continuous spectra do not follow the KZ law. Well-known examples are the elastic thin vibrating steel plate, \cite{Mor08}, and gravity waves in mercury, \cite{fauve}. Also, surface waver waves produced in laboratory in a flume of size $12\times 6\times1.5$ m developed a strongly non-linear discrete energy cascade lacking any continuous spectrum, \cite{denis}. The form of the energy spectra depends, in addition, on the parameters of the initial excitation, \cite{LNMD09,XiSP10}. In experiments with capillary waves in He strong non-locality was found; the local wave-amplitude maximum appeared at frequencies $\omega$ of the order of the viscous cut-off, \cite{ABKL09}.

Attempts on improving the statistical wave turbulence theory  either refer to frozen \cite{Pu99}, sandpile \cite{Naz06}, mesoscopic \cite{ZKPD05} or finite-dimensional  models \cite{LPPR09}, neither allowing for the inclusion of  the dependence on the initial conditions or finite sizes. In the present Letter we propose a novel model (called further on D-model) based on  discrete wave turbulence \cite{K-Discrete, CUP} and including energy cascades. It provides a uniform frame for studying finite-size effects and deducing the generic form of turbulent energy spectra depending on the parameters of initial state.\\

\textbf{2. D-model.}

\noindent \textbf{\emph{2.1. Excitation of a (quasi-) resonant mode.}}

2.1.1. Within the discrete model of wave turbulence [16] the solutions of the resonance conditions form a group of independent resonance clusters, each cluster being graphically represented by its non-linear resonance diagram (NR-diagram) which explicitly defines the dynamical system of the cluster and its conservation laws. The solutions of the corresponding dynamical system may be either regular or chaotic, depending on the structure of the cluster and the details of the initial excitation.  This holds for quasi-resonance with small frequency mismatches.

2.1.2. (Quasi-) resonant modes  occur all-over k-space; they are not restricted to the range $k<k_1$.

\begin{figure}
\begin{center}\vskip -0.1cm
\includegraphics[width=5cm]{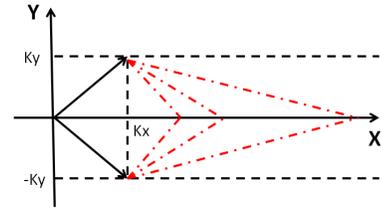}
\caption{\label{f:nonloc} Color online. Nonlocal interactions in a 4-wave system, $\o \sim k^{\a}$. Each couple of (red) dashed lines with equal lengths correspond to specific choice of a parameter $t$.}
\end{center}\vskip -0.7cm
\end{figure}

2.1.3. (Quasi-) resonance interactions may not be localized in $k$-space. In a 4-wave system with dispersion function $\o \sim k^{\a}$ modes with \emph{arbitrary large wavelength difference} can interact directly (see Fig.\ref{f:nonloc}). In this case a parametric series of solutions of resonance conditions
$
k_1^{\a}+k_2^{\a} =k_3^{\a}+k_4^{\a}, \ \
 \vec k_1+\vec k_2=\vec k_3+\vec k_4, $
can be easily written out as
$
\vec k_1= (k_x,k_y), \ \ \vec k_2=(\mathbf{t},-k_y), \ \ \vec k_3=(k_x,-k_y), \ \ \vec k_4=(\mathbf{t},k_y),
$
with real parameter $t$.

2.1.4. In addition, non-resonant modes, i.e. modes of large frequency mismatch, do exist. In fact, most of the modes in a 3-wave system are indeed non-resonant. These modes do not change their energies on the corresponding resonant time-scale.

\noindent \emph{\textbf{2.2. Excitation of a non-resonant mode.}}

In this case the generic mechanism of  instability  can be described as an interaction of three monochromatic wave trains $\o_+ + \o_- = 2\o_0$ where
\be \label{Phil}
\o_+=\o_0 + \Delta \o, \,\, \o_-=\o_0 - \Delta \o, \,\, 0<\Delta \o<1.
\ee
This type of instability is quite general and is known in various areas of physics under different names, e.g. parametric instability in classical mechanics, Suhl instability  of spin waves, Oraevsky-Sagdeev  decay instability of plasma waves, modulation instability in nonlinear optics, Benjamin-Feir instability in deep water, etc., \cite{ZO08}.

Conditions for the modulation instability to occur have been written in terms of the increment of instability $I$;
its form may differ for diverse wave systems and for the same wave system with different magnitudes of the nonlinearity parameter (e.g. \cite{Be63} and \cite{DY79}, for surface water waves). However, as $I$ can always be presented as a polynomial on $\o, \, k$ and $\Delta \o$, it does not affect our general scheme (presented below) for computing energy spectra. To shorten further computation we have chosen the simplest form of the instability increment according to \cite{Be63} (weakly nonlinear Schr\"{o}dinger equation with the value of the nonlinearity parameter $\varepsilon\sim 0.1$ to 0.2):
\be \label{BFI-incr}
0< I=\frac{\Delta \o}{\o A k} < 1.
\ee
Aiming to construct a stationary ("saturated") energy spectrum
we regard  cascading chain of the form
\bea \label{general}
\begin{cases}
2\o_{0}= \o_{1,1}+\o_{2,1},    &E_1=p_1E_0, \, , \\
\o_{1,1}=\o_{2,1}+\o_{2,2},  &E_2=p_2E_1, \\
....  \\
 \o_{n-1,1}=\o_{n,1}+ \o_{n,2},   &E_n =p_n E_{n-1}
\end{cases}
\eea
where $0< p_j < 1,$ $\o_f$ is the forcing frequency, $E_j$ is the energy  at the $j$-th step of  cascade and $p_j$ is the part of the energy $E_{j-1}$ transported  from cascading mode $A_{j-1}$ to cascading mode $A_j$. Each cascade step is regarded independently at discrete time moments.

\emph{2.2.1. Assumptions.} We assume further  that

(1) $p_j=p=\const$, i.e. \emph{cascade intensity} $p$ is constant for given excitation parameters; in this case $p=(1+\sqrt{1-E_0})/2$, where the energy of the initial excitation is normalized to 1 and $E_{0}$ is the energy of its stationary amplitude $A_0.$

(2) at each cascade step $j$, one cascading mode with frequency $\o_{j,1}$ is generated, with maximal increment of instability
\vskip -0.6cm
\be \label{BFI-incr-max}
I_{\mathbf{max},n}=\frac{|(\Delta \o)_n|}{\o_n A_n k_n}= 1.
\ee
(this corresponds to the idea of Phillips who suggested that in the saturated range the spectral density is saturated at a level determined by wave breaking). Modes $\o_{j,2}$ have smaller increments of instability and are fluxless ("frozen", \cite{Pu99}).

\emph{2.2.2. The chain equation.} Cascading chain  consists of modes with frequencies $\o_{1,1}, \o_{2,1}, \o_{3,1},...$; it can be shown that the distance between two cascading modes $|(\Delta \o)_n|$ decreases slowly with growing $n$. It follows from (1) that
\be \label{p}
E_n =p^n E_0 \RA A_{n+1}=\sqrt{p} A_n,
\ee
The second assumption (2) can be rewritten as
\be \label{inc-n1}
 \o_{n+1}=\o_n \pm  \o_n A_n k_n,
\ee
and combination of $A_{n+1}=\sqrt{p} A_n$ and (\ref{inc-n1}) yields
\be \label{1}
\sqrt{p} A_n= A(\o_n \pm \o_n A_n k_n)
\ee
which is further on called \emph{the chain equation} ("+" and "-" stand for direct and inverse cascade, accordingly).
\begin{figure*}
\vskip -0.1cm
\begin{center}
\includegraphics[width=7cm,height=4cm]{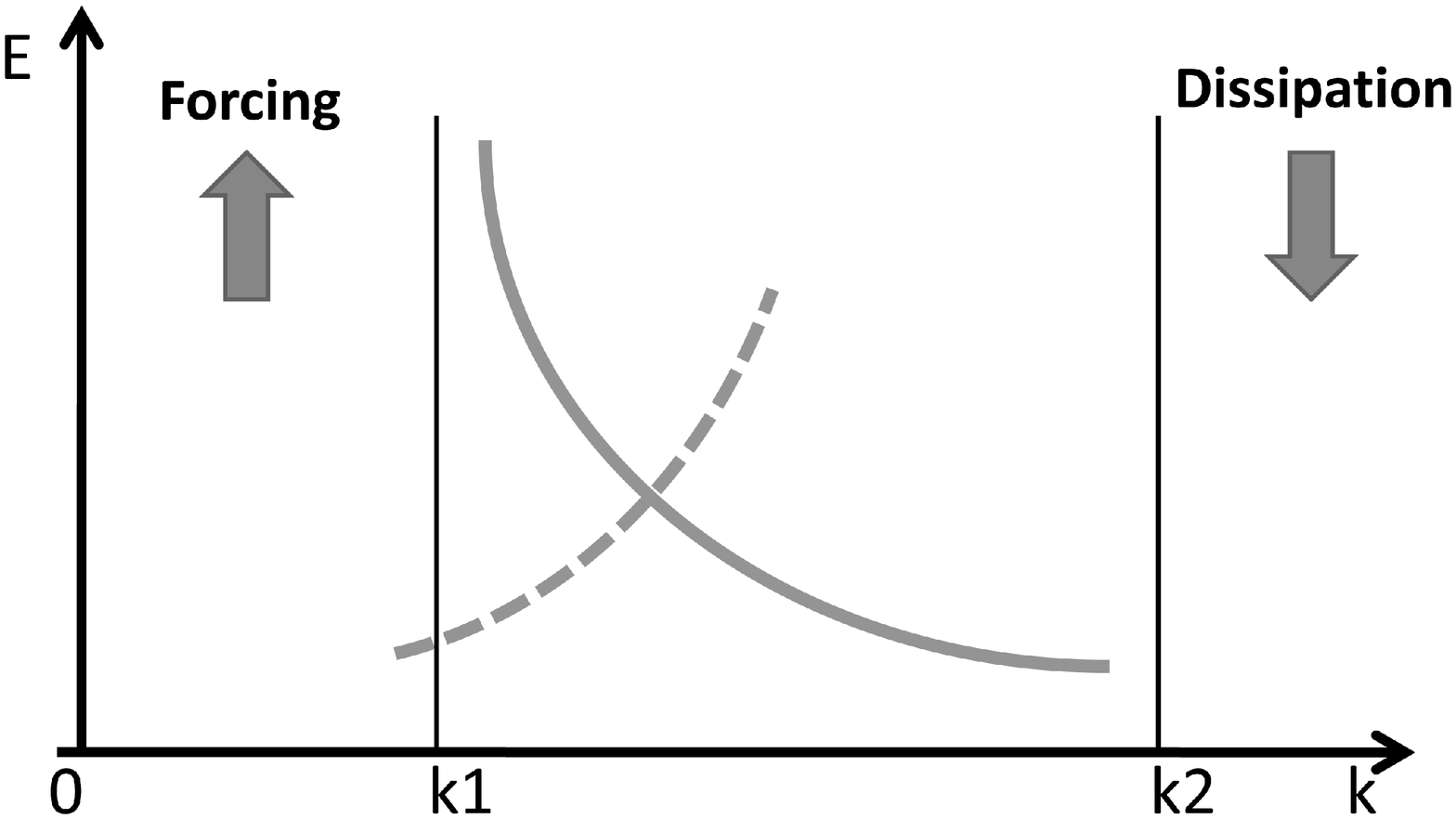}
\includegraphics[width=7.8cm,height=4cm]{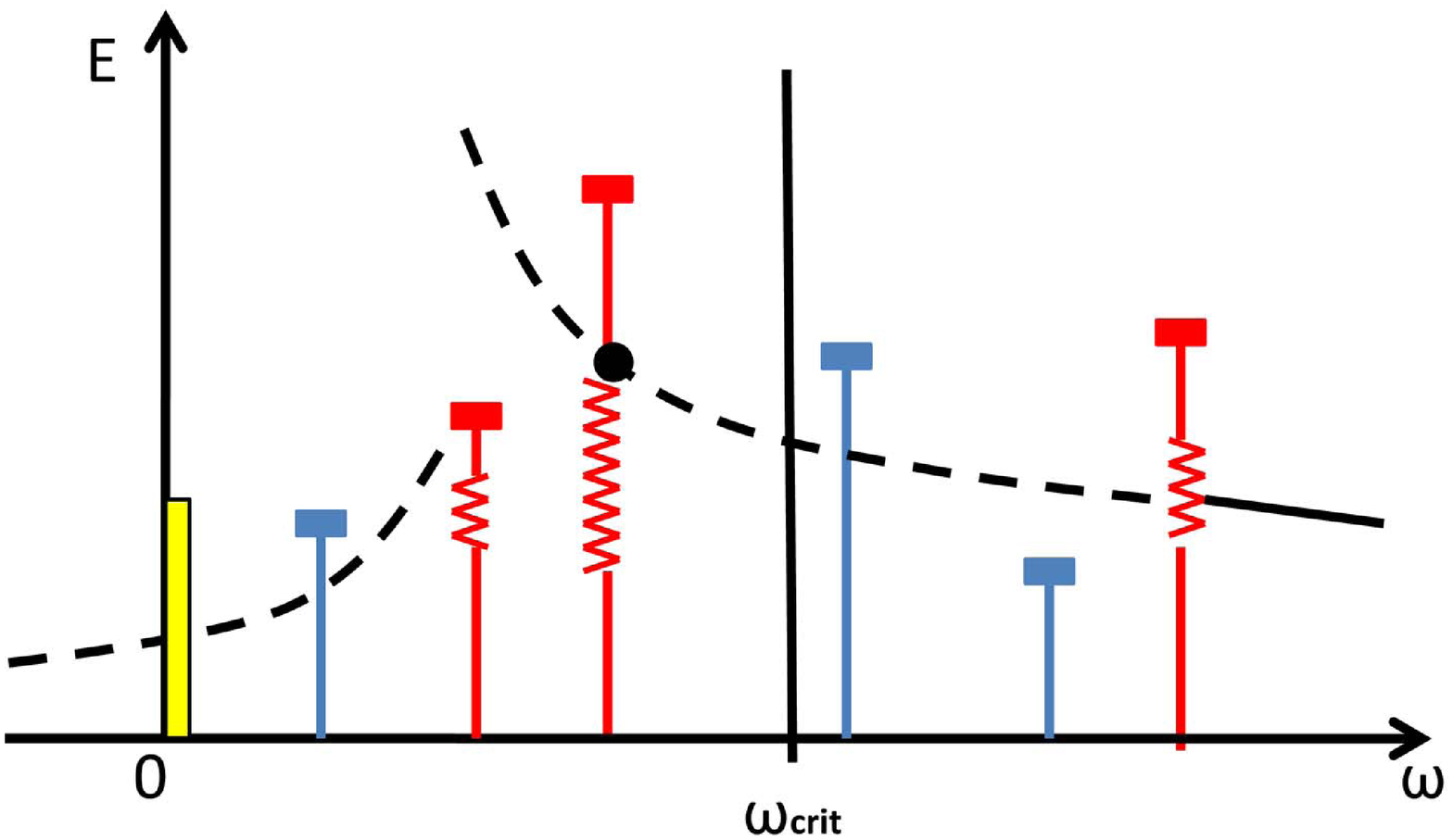}
\caption{\label{f} Color online. Classical model of WT (on the left) and D-model of WT (on the right).}
\end{center}
\vskip -0.7cm
\end{figure*}

\emph{2.2.3. Energy spectra.} Taking Taylor expansion of  RHS of the chain equation yields
\bea \label{Taylor}
\sqrt{p} A_n= A(\o_n \pm \o_n A_n k_n)= \sum_{s=0} ^ {\infty } \frac {A_n^{(s)}}{s!} \, (\pm\o_n A_n k_n)^{s} \nonumber \\
 = A_n \pm  A_n^{'}\o_n A_n k_n+ \frac 12 A_n^{''}(\pm \o_n A_n k_n)^2+...
\eea
Taking two first RHS terms from  (\ref{Taylor}) we get
\be\label{2terms}
\sqrt{p} A_n = A_n \pm  A_n^{'}\o_n A_n k_n
\RA A_n^{'} = \pm \frac{\sqrt{p}-1}{\o_n k_n } \RA   \ee
\be \label{A-gen}
A(\o) = \pm (\sqrt{p}-1) \int \frac{d \o}{\o k}+\mathbf{const}(\o_0, A_0)
\ee
Substitution of some specific dispersion relation into (\ref{2terms}) yields dependence $A=A(\o)$ and the form of energy spectrum, $E \sim A^2$. E.g. if  $\o=k^{\a}, \, \a >0,$ (\ref{A-gen}) yields
\bea
A(\o) &=& \mp \frac{(1-\sqrt{p})}{2+\a^{-1}} (\o^{-(2+\a^{-1})}-\o_0^{-(2+\a^{-1})})+A_0, \nn \\
E(\o) &\sim& \o^{-\nu} \quad \mbox{with} \quad  2+\a^{-1}\le \nu \le 2(2+\a^{-1}). \nn
\eea
This is in accordance with the laboratory results reported e.g. in \cite{denis,LNMD09}, for $\o \sim k^{1/2}$.

\emph{2.2.4. Direction and termination of  cascades.} Conditions for  cascade's termination and for  formation of the direct and inverse cascades can be easily obtained from the form of the increment (\ref{inc-n1}). Namely,
cascade terminates at some step $N$  if (\ref{BFI-incr}) is violated:
\be \label{termin}
\o_{N+1}-\o_N =0 \quad \mbox{or}\quad |\o_{N+1}-\o_N|> \o_N A_N k_N.
\ee
Direct and inverse cascades occur if
\be \label{cas-dir-inv}
\o_{n+1}-\o_n >0 \quad \mbox{or}\quad \o_n - \o_{n+1} >0
\ee
accordingly. Depending on the value of excitation parameters, $N< \infty$ or $N= \infty$, \cite{KSh11-2}.

A simplified form of  cascade with $|(\Delta \o)_n|=\const$, \cite{KSh11}, though useful for interpreting some experimental results,  yields cascade's termination at the final step and does not give a lead for  possible transition to  continuous spectrum.\\

\textbf{3. Classical model \emph{versus} D-model.}

\noindent Graphical presentation of the classical model  and D-model of WT is shown in Fig.\ref{f} (in the left and right panels correspondingly). One of  crucial assumptions of the classical model is, that forcing and dissipation are  wide apart in the k-space, while the energy transport described by stationary spectra takes place within inertial interval $(k_1,k_2)$. This assumption is very difficult to reproduce either in laboratory experiments or numerical simulations.
"Finite-size" effects are supposed to appear in the range of wavelengths $k< k_1$ and are not described by the classical model, while in the range $k>k_2$ energy transport is supposed to be suppressed by dissipation. This phenomenology does not depend on the details of the initial forcing.

On the other hand, in the D-model  dissipation might be switched on or off at any cascade step by changing the magnitude of  cascade intensity $p$, i.e. no additional assumptions are needed about relative positions of forcing and dissipation in the k-space; the existence of  inertial interval is not important either and  initial excitation may take place all over the k-space. In this model, three types of modes are singled out; energy transport in  wave system depends on the type of the initially excited mode.\\

\emph{3.1. Resonant and quasi-resonant} modes (shown as red vertical solid T-shaped lines having a spring-like part); being excited, they may yield chaotic or recurrent patterns; no cascade shall be generated (e.g. \cite{HHS03}). In this case \emph{time synchronization} condition (for frequencies) is satisfied exactly or with a small frequency mismatch while \emph{space synchronization} condition (for wave vectors) is satisfied exactly:
\bea \o_1+\o_2&=&\o_3 +\Omega, \ \vec{k}_1+\vec{k}_2=\vec{k}_3, \ 0 \le \Omega \ll 1  \ \mbox{or} \nn\\
\o_1+\o_2&=&\o_3+\o_4+\Omega, \ \vec{k}_1+\vec{k}_2=\vec{k}_3+\vec{k}_4, \ \nn\eea
and phases are coherent.\\

\emph{3.2. Cascading modes} (shown as small dashes composing black dashed curves); being excited, they form a cascading chain with energy discrete spectrum $\sim A^2$ with $A$ computed from (\ref{A-gen}); for some specific initial magnitudes $A_0$ and $\o_0$) energy spectrum might turn into continuous (shown as black solid "tail"  of the curve) though not necessarily a classical spectrum. Choice of the parameters of initial excitation also defines whether excitation of a cascading mode yields direct or inverse cascade (e.g. \cite{TW99}). In this case \emph{the time synchronization} condition (\ref{Phil}) is satisfied exactly, while \emph{the space synchronization} condition is violated:
\be
\o_1+\o_2=2\o_3, \ \vec{k}_1+\vec{k}_2=2\vec{k}_3+ \Theta, \ 0 \le \Theta \ll 1, \nn\ee
and phases are are not coherent.\\

\emph{3.3. Frozen modes} (shown as blue vertical solid T-shaped lines): these are fluxless modes which do not take part in  energy transport over the k-space, they may occur for some choice of initial state (e.g. \cite{Pu99,PRL}). For frozen modes \emph{both time and space synchronization conditions are violated}.
This means in particular that fluxless modes
$\o_{1,2}, \o_{2,2}, \o_{3,2},...$ have non-coherent phases and provide a necessary prerequisite in the statistical theory for a possible further development of a classical spectra. If at some cascade step a (quasi-) resonant mode is generated (shown as a black circle, Fig.\ref{f}, on the left) a cascading chain may be broken by  appearance of an intermittency.\\

It is important to realize that increment of instability is a parameter characterizing interplay of dispersion and nonlinearity in a wave turbulent system. Accordingly,
two conditions of  cascade termination (\ref{termin}) have different physical consequences:
$\o_{N+1}-\o_N =0$ yields $I=0$ and manifests  transition to a linear regime, while $|\o_{N+1}-\o_N|> \o_N A_N k_N$ corresponds to the case $I>1$ and growing nonlinearity. For instance, it has been shown in \cite{KSh11-2} for surface water waves that starting with a small nonlinearity $\varepsilon_0 =A_0\o_0=0.1$, initial amplitude looses an order in the  magnitude after 31 steps of the cascading chains while the steepness of a wave packet becomes 0.288. This means that another form of the increment should be used further on, e.g.
\vskip -0.5cm
\be \label{Dys}
I=|\Delta \o|/\Big(\o  A k - \frac{3}{2}\o^2  A^2 k\Big),
\ee
\vskip -0.3cm
obtained in \cite{DY79} for weakly nonlinear Schr\"{o}dinger equation with parameter of nonlinearity $\varepsilon \sim 0.25$ to 0.4. This means also that the condition $|\o_{N+1}-\o_N|> \o_N A_N k_N$ allows to compute a critical frequency $\o_{crit}$ (shown as a (black) bold vertical line at Fig.\label{f:fig1}, on the right) at which nonlinearity becomes mode important that dispersion. The energy spectra $E(\o), \, \o>\o_{crit}$ would differ substantially from the classical spectra.

Another interesting phenomenon observed in numerous laboratory experiments is formation of a narrow zero-frequency sideband with non-zero energy (shown as vertical yellow rectangular). It should be checked whether this phenomenon might be attributed to termination of an inverse cascade or originates in the (quasi) resonant interactions of short eigen-modes (appearance of an intermittency).\\

\textbf{4. Conclusions.}

\noindent In this Letter our main goal was to
 understand natural phenomena revealed by laboratory observations of WT and to find the physical mechanism that underlies it. We have developed a simple model and analyzed it to see if it captures critical aspects of the phenomenon.

 We have demonstrated that D-model of WT allows to reproduce a number of discrete effects including generation of a set of cascading modes in a wave turbulent system. We have also deduced the corresponding energy spectrum depending on the form of dispersion relation $\o=\o(k)$, the form of the instability increment and parameters of the initial excitation. Conditions for formation of a direct or an inverse cascade have been written out explicitly. It has been shown that, depending on  initial state, a cascade might yield  decrease or growth of the nonlinearity.

A few possible ramifications of the model are accounted for below (the list is not exhaustive):

\noindent\textbullet~to change the form of the instability increment (\ref{BFI-incr}), e.g. to (\ref{Dys}); a number of examples is given in \cite{ZO08};

\noindent\textbullet~to regard cascade intensity as a function of wavelengths, $p=p(k)\neq \mathbf{const}$, then
 ODE similar to (\ref{2terms}) is also solvable in quadratures:
\vskip -0.7cm
 \be \label{A-gen-dis}
A(\o) = \pm  \int \frac{(\sqrt{p(k)}-1) d \o}{\o k}+\mathbf{const}(\o_0, A_0).
\ee
\vskip -0.3cm
For instance,  power function $p(k_j)=k_j^{\g}$ may model dissipation;  piecewise constant function $p=p_j, \forall k\in (k_{j-1}, k_j)$  may model wind-generated surface water waves at different fetches $(k_{j-1}, k_j)$, etc.

\noindent\textbullet~to regard the hierarchy of D-models with refined form of turbulent energy spectra obtained by cutting the Taylor expansion (\ref{Taylor}) at 3, 4 and so on terms instead of the simplest equation (\ref{2terms}).  It would be a challenge to find the general solution of the infinite order ODE given by (\ref{Taylor}) and to investigate  bounds of its physical applicability (it should be verified that energy flux corresponding to  chosen ODE is not linear).

{\textbf{Acknowledgements.}} Author acknowledges
 M. Shats, I. Shugan and R. Treumann for valuable discussions. This research has been supported by
the Austrian Science Foundation (FWF) under project
P22943-N18 "Nonlinear resonances of water waves" and in part by the National Science Foundation, USA, under Grant No. NSF PHYS05-51164.

\end{document}